\documentstyle[preprint,aps]{revtex}
\begin{document}
\tighten
\title{On the possibility of an $\eta$--meson light nucleus bound
state formation}
\author{
S. A. Rakityansky\footnote{Permanent address:
Joint Institute  for Nuclear Research, Dubna, 141980, Russia},
S. A. Sofianos,}
\address{ Physics Department, University of South Africa,
 P.O.Box 392, Pretoria 0001, South Africa}
\author{V. B. Belyaev,}
\address{ Joint Institute  for Nuclear Research, Dubna, 141980,
Russia}
\author{W. Sandhas}
\address{
 Physikalisches Institut, Universit\H{a}t Bonn, D-53115 Bonn, Germany}
\date{\today}
\maketitle

\begin{abstract}
The resonance and bound--state poles of the amplitude describing elastic
scattering of $\eta$-meson off the light nuclei $^2H$,\,$^3H$,\, $^3He$, and
$^4He$ are calculated in the framework of a microscopic approach
based on few--body equations. For each of the nuclei, the two--body
parameters that enhance the $\eta N$--attraction which generate
quasi--bound states, are also determined.
\end{abstract}
\vspace{.5cm}

\section{Introduction}
Experiments on production of low-energy $\eta$--mesons in nucleus--nucleus
\cite{fras}, $N$--nucleus \cite{ber,chi}, and $\pi$--nucleus \cite{peng}
collisions, revealed an energy dependence of the corresponding
cross--sections, which does not contradict the expected strong
$\eta$--nucleus final--state interaction \cite{wilkin}. This is natural,
since the low--energy $\eta N$--~interaction is resonant. The possibility
that this energy dependence is related to the existence of an $\eta$--nucleus
quasi--bound state cannot be excluded. Indeed it was shown in Ref.\cite{bhal}
that at low energies the $\eta N$ interaction, in the dominant $S_{11}$
channel  where the near--threshold resonance  $N^*(1535)$ is excited,
is attractive.\\

On the other hand estimations obtained in the framework of the
first-order optical potential theory \cite{hai,liu}, put a lower bound on
the nucleus atomic number $A$ for which an $\eta$--nucleus bound state could
exist, namely, $A\ge12$.\\

However, some speculations on the possibility of a formation of
$\eta$--helium bound state still
appear\cite{wyc} despite the discouraging results of the first ( and the
sole ) experiment \cite{chri} on direct search for bound states of
$\eta$--meson with Lithium, Carbon, Oxygen, and Aluminium. All such
speculations are based on the large negative values ( $\sim-2$ fm ) of the
real parts of $\eta$--nucleus scattering lengths calculated within a
simplified optical--potential theory \cite{wilkin,wyc}.\\

To the best of our knowledge, the only microscopic calculations of the
scattering lengths was presented in our recent papers \cite{fid,rak,bel}.
In these, it turned out that the $\eta$--helium scattering length
could have even larger ( negative ) real part than earlier estimations. This
raises more doubts on the validity of the abovementioned constraint, $A\ge12$,
for the existence of $\eta$--nucleus bound states.\\

In the present work we examine the possibility of formation of a bound state
in the  $\eta$--meson  $d$, $t$, $^3\!He$, and $^4\!He$ systems, in the
framework of a microscopic approach, namely, the Finite-Rank Approximation
( FRA ) of the nuclear Hamiltonian \cite{bel1,bel2}. The approximate few--body
equations of this approach enable us to calculate the $\eta$--nucleus
$T$--matrix $T(\vec k',\vec k;z)$ for any complex total energy
$z$, i.e. for any point on the complex plane of the momentum
$p=\sqrt{2\mu z}$. In this way we have numerically located the resonance
state poles and (for each of the nuclei) we found the parameters that
enhance the $\eta N$--attraction which generate quasi--bound state.\\

\section{The method}
The total Hamiltonian of the quantum system consisting of an $\eta$--meson
and a nucleus of atomic number $A$,  can be written as
\begin{equation}
	  H = H_0 + V + H_A ,
\end{equation}
where $H_0$ is the $\eta$ - nucleus  kinetic energy operator (free
Hamiltonian), $ V = V_1 + V_2 + \cdots+ V_A $   is the sum of
$\eta N$--potentials, and $H_A$  is the total Hamiltonian of the nucleus.
Elastic scattering process is the transition from  the initial
 $| \vec {k},\psi_0 >$ to the final $|\vec{ k}',\psi_0 >$  asymptotic state
which differ only in the direction of the relative $\eta$--nucleus
momenta $\vec k$ and $\vec k'$. During this process,  the nucleus  remains
in the ground eigenstate $|\psi_0 >$,  of $H_A$
\begin{equation}
   H_A |\psi_0 >  =  {\cal E}_0 |\psi_0 >.
\end{equation}
The $\eta$-nucleus scattering amplitude $f(\vec{k}',\vec{k};z)$,
is expressed in terms of the asymptotic states
\begin{equation}
       f(\vec{k}',\vec{k};z)=-\frac{\mu}{2\pi}\,
      <\vec{k}',\psi_0|T(z)|\vec{k},\psi_0>,
\end{equation}
i.e in terms of the matrix elements
of the operator $T$ obeying the Lippmann-Schwinger equation
\begin{equation}
\label{T}
       T(z)=V+V\frac{1}{z-H_0-H_A}T(z).
\end{equation}
Here $\mu$ is the $\eta$--nucleus reduced mass.
For further developments, we introduce the (auxiliary) operator $T^0$ via
\begin{equation}
\label{T0}
         T^0(z)=V+V\frac{1}{z-H_0}T^0(z)
\end{equation}
and rewrite Eq. (\ref{T}) in the form
\begin{equation}
\label{Top}
       T(z)=T^0(z)+T^0(z)\frac{1}{z-H_0} H_A
               \frac{1}{z-H_0-H_A} T(z).
\end{equation}
Then, using the approximation
\begin{equation}
\label{c}
     H_A\approx {\cal E}_0|\psi_0><\psi_0| \label{Happrox}\,,
\end{equation}
we obtain  for the matrix elements
$
  T({\vec k}',{\vec k};z)\equiv <\vec{k}',\psi_0|T(z)|\vec{k},\psi_0>
$
the following integral equation
\begin{equation}
\label{tm3}
   T(\vec{k}',\vec{k};z) = <\vec{k}',\psi_0|T^0(z)|\vec{k},\psi_0>
    +  {\cal E}_0\int \frac{d^3k''}{(2\pi)^3}
        \frac{ <\vec{k}',\psi_0|T^0(z)|\vec{k}'',\psi_0>}{
       (z-\frac{{k''}^2}{2\mu})(z-
        {\cal E}_0-\frac{{k''}^2}{2\mu})}\,T(\vec{k}'',\vec{k};z),
\end{equation}
which after the partial--wave decomposition becomes one-dimensional and can
be easily solved (numerically) if we know the auxiliary matrix
\begin{equation}
 <\vec{k}',\psi_0|T^0(z)|\vec{k},\psi_0>.
\end{equation}
It is easy to see that the $T^0$--operator describes the scattering
 of an $\eta$--meson from  nucleons fixed at their spatial
 positions inside the nucleus, because the equation (\ref{T0}), which defines
$T^0$,  does not contain any operator which acts on the
internal nuclear Jacobi coordinates $\{\vec{r}\}$. Therefore all operators
 in Eq. (\ref{T0}), are  diagonal in these variables and thus
its momentum representation reads
\begin{equation}
            T^0(\vec{k}',\vec{k};\vec{r};z)=V(\vec{k}',\vec{k};\vec{r})
                + \int \,\frac{d^3k''}{(2\pi)^3}
                 \frac{V(\vec{k}',\vec{k}'';
                 \vec{r})}{z-\frac{k''^2}{2\mu}}
                 \,T^0(\vec{k}'',\vec{k};\vec{r};z).
\label{t0m}
\end{equation}

\noindent
In other words it depends only parametrically on the
coordinates $\{\vec{r}\}$. Solving this integral equation and
employing the ground state wave function  $\psi_0(\vec r)$ for the
nucleus, the input $<\vec{k}^\prime, \psi_0 | T^0(z)|\vec{k}, \psi_0>$ to
Eq. (\ref{tm3}) is obtained by  integrating over all nuclear Jacobi
coordinates $\{\vec r\}$
\begin{equation}
    <\vec{k}^\prime, \psi_0 | T^0(z) | \vec{k}, \psi _0 > =
     \int d^{3(A-1)}r |
      \psi_0(\vec{r})|^2 T^0(\vec{k}^\prime,\vec{k};\vec{r};z).
\label{aver}
\end{equation}
For practical calculations, Eq. (\ref{T0})  is written in terms of the
Faddeev components $T^0_i(z)$
\begin{equation}
T^0(z)= \sum^A_{i=1} T^0_i(z)\,,
\end{equation}
Introducing the operators
\begin{equation}
     t_i(z)=V_i+V_i\frac{1}{z-H_0}t_i(z),
\end{equation}
we finally  get for  $T^0_i$  the following system
of integral equations
\begin{eqnarray}
     T^0_i (\vec{k}',\vec{k};\vec{r};z)
     &=& t_i(\vec{k}',\vec{k};\vec{r};z) \
  + \int \frac {d^3k''}{(2\pi)^3}
\frac {t_i(\vec{k}',\vec{k}'';\vec{r};z)} {z - \frac {k''^2}{2\mu}}
\sum_{j\neq i} T^0_j(\vec{k}'',\vec{k};\vec{r};z) .
\label{t0i}
\end{eqnarray}
\noindent
The $t_i$ describes the scattering of the $\eta$-meson off the
$i$-th  nucleon. It is expressed in terms of the corresponding two-body
$t_{\eta N}$-matrix via
$$
t_i(\vec k',\vec k;\vec r;z)=t_{\eta N}(\vec k',\vec k;z)
 \exp\left[{{\displaystyle i(\vec k-\vec k')\cdot\vec r_i}}\right]
$$
where $\vec r_i$ is the  vector from the nuclear center of mass to the $i$-th
nucleon and can be expressed in terms of the Jacobi vectors
$\{\vec r\}$.\\

We emphasize, that the use of $T^0$ in the above scheme is not an additional
(fixed--scatterer) approximation. The coupled equations (\ref{T0}) and
(\ref{Top}) are exact. The only approximation used here is the one defined by
Eq. (\ref{c}) and corresponds to a truncation of the spectral expansion of
the nuclear Hamiltonian. Physically, it means that during the multiple
scattering of $\eta$- meson, the nucleus remains in its ground state. This
approximation is widely used in nuclear physics and is known as the coherent
approximation \cite{Kerm}.
\section{Results and discussion}

As an input information  we need  the ground-state wave
functions $\psi_0$ of the nuclei involved and the two-body $t$--matrix
$t_{\eta N}$.

For the bound states  we employed simple Gaussian--type functions
\begin{eqnarray}
\psi_d(\vec{{\tt x}}) & = & \left ( \frac {3}{8\pi < r^2_d >} \right )^{3/4}
\exp \left ( - \frac {3{\tt x}^2}{8 < r^2_d>} \right ),  \\
\psi_t(\vec{{\tt x}}, \vec{{\tt y}}) &= &(\sqrt{3} \pi < r^2_t >)^{-3/2}
\exp \left [ - \left ( \frac {{\tt x}^2}{2} + \frac {2{\tt y}^2}{3}
\right )/(2 < r^2_t >) \right ], \\
\psi_{\alpha} (\vec{{\tt x}}, \vec{{\tt y}}, \vec{{\tt z}})&=&\left [ 2
   \left ( \frac {9}{16 \pi <r^2_d>} \right )^3 \right ]^{3/4}\exp
   \left [ - \frac {9}{16<r^2_\alpha>} \left ( \frac {{\tt x}^2}{2} + {\tt y}
^2 + \frac {{\tt z}^2}{2} \right ) \right ],
\end{eqnarray}
where $\vec{{\tt x}}, \vec{{\tt y}},$ and $\vec{{\tt z}}$ are the Jacobi
vectors . These functions were constructed to be symmetric
with respect to nucleon permutations, and to reproduce the experimental
mean square radii:
          $\sqrt{<r_d^2>}$=1.956 fm \cite{rmsD},
          $\sqrt{<r_{{^3H}}^2>}$=1.755 fm \cite{rmsT},
          $\sqrt{<r_{{^3He}}^2>}$=1.959 fm \cite{rmsT}, and
	  $\sqrt{<r_{\alpha}^2>}$=1.671 fm \cite{Til4}.
For masses and binding energies of the nuclei we  used the
experimental values  \cite{Waps}.

Since at low energies the $\eta N$ interaction is dominated by
the $N^*(1535)$   $S_{11}$ - resonance, we used the following separable
form  for the $\eta N$ - amplitude
\begin{equation}
           t_{\eta N}(k',k;z) = \frac {\lambda}{(k'^2+
           \alpha^2)(z - E_0 + i\Gamma/2)(k^2+\alpha^2)}
\end{equation}
with $E_0 = 1535\; {\rm MeV} - ( m_N + m_{\eta} )$  and $\Gamma = 150\;
{\rm MeV}\;$ \cite{PDGr}. In order to fix the parameter $\alpha$,
we make use of the results of Refs. \cite{bhal,Benn}, where the
same $\eta$N $\to$ N$^*$ vertex function ($k^2 + \alpha^2)^{-1}$ was
employed with $\alpha$ being determined via a two-channel fit to the
$\pi $N $\to \pi $N and $\pi $ N $\to \eta $N experimental data. \\

Due to experimental uncertainties and differences between the models for the
physical processes, one can use three different values for the  range
parameter $\alpha$, namely, $\alpha = 2.357$~fm$^{-1}$ \cite{bhal}, \,\,
$\alpha = 3.316$~fm$^{-1}$ \cite{Benn},\, and $\alpha = 7.617$~fm$^{-1} $
\cite{bhal}. Since there is no criterium for singling out
one of them, we use all three in our calculation.\\

The remaining parameter $\lambda$ is chosen   to provide the correct
zero-energy on-shell limit, i.e., to reproduce the known $\eta N$
scattering length $a_{\eta N}$,
\begin{equation}
 t_{\eta  N}(0,0,0) = - \frac {2\pi}{\mu_{\eta  N}}a_{\eta  N}.
\end{equation}
Like the range parameter $\alpha$, the scattering length $a_{\eta N}$ is
not well known, the estimated values being within the range $Re\,a_{\eta N}\in
 [0.27,0.98]\,
{\rm fm}$ and $Im\,a_{\eta N}\in [0.19,0.37]\,{\rm fm}$ \cite{serb}.
Our intention is to vary the $Re\,a_{\eta N}$ until the corresponding
$\eta N$ attraction generates a bound state in the  $\eta$--nucleus system.
As the starting value we chose  $(0.55+i0.30)$ fm proposed by Wilkin
\cite{wilkin}. Thus, we take
\begin{equation}
 \label{length}
       a_{\eta N}=(g0.55+i0.30)\,{\rm fm},
\end{equation}
where $g$ is an enhancing parameter.\\

Since $a_{\eta N}$ is complex, the $\eta$--nucleus Hamiltonian is
non--Hermitian and its eigenenergies are generally complex. Hence, we do not
expect to find a pole of $T(z)$ on the positive imaginary axis of the
complex k--plane with any choice of the enchancing factor $g$. As was shown
in Ref.\cite{cass}, when the interaction becomes complex the bound-state poles
move into the second quadrant of the complex k--plane. Therefore we search
in this quadrant in order to locate  possible  poles of $T(z)$.
However, not all poles in the  second quadrant stems from bound states.
Indeed, the energy $E_0=p^2_0/2\mu$ corresponding to a pole,
\begin{equation}
E_0=\frac{1}{2\mu}\left[(Re\,p_0)^2-(Im\,p_0)^2+2i(Re\,p_0)(Im\,p_0)\right]\,,
\end{equation}
has a negative real part only if $p_0$ is above the diagonal of this
quadrant. Below the diagonal $Re\,E_0>0$ the pole is attributed
to a resonance. Therefore this diagonal is the critical border, and when
crossing from below,  a pole becomes a quasi--bound
state.\\

Fixing the enhancing factor $g$ of Eq. (\ref{length}) to the value
$g=1$ and making variations of the complex parameter $p=\sqrt{2\mu z}$
within the second quadrant, we located the
poles close to the origin, $p=0$,  which are given in the Table 1.\\

It is seen, that for the $\eta d$, $\eta t$, and $\eta{^3\!He}$ systems,
these poles lie below the diagonal, i.e. in the resonance
region, while the $\eta{^4\!He}$ system has a quasi--bound state.
On the other hand, all such poles are not far from the border separating
the resonance and bound state domains, since the difference between
$|Re\,p_0|$ and $|Im\,p_0|$ is not very large and the $|Re\,E_0|$ is rather
small for all poles found.  Hence, one can expect that
small changes of the factor $g$ could place the poles on this border.
Following this idea, we varied $g$  untill we  found the factors
which generates poles on the diagonal. They are given in the Table 2. \\\\
Table 1.\,\,
Positions $p_0=\sqrt{2\mu E_0}$ of the poles of the $\eta$--nucleus
amplitudes with $g=1$.
For each of the nuclei the calculations were done with three
values of the range parameter $\alpha$.
\begin{center}
\begin{tabular}{|c|c|c|c|}
\hline
 & $p_0\;({\rm fm}^{-1})_{\mathstrut}^{\mathstrut}$ &
             $E_0\;({\rm MeV})$ &
             $\phantom{-}\alpha\;({\rm fm}^{-1})\phantom{-}$\\
\hline\hline
    & $-0.90254+i0.35880$ & $ 31.448-i29.698 $ & 2.357 \\
\cline{2-4}
$\eta\, d$ & $-0.84562+i0.32422$ & $ 27.969 -i25.143 $ & 3.316 \\
\cline{2-4}
    & $-0.82460+i0.30855$ & $ 26.813 -i23.333 $ & 7.617 \\
\hline\hline
    & $-0.56125+i0.24475$ & $ 10.818 -i11.650 $ & 2.357 \\
\cline{2-4}
$\eta\, t$ & $-0.55747+i0.27050$ & $ 10.076 -i12.789 $ & 3.316 \\
\cline{2-4}
    & $-0.52717+i0.28349$ & $ 8.3770 -i12.675 $ & 7.617 \\
\hline\hline
    & $-0.54791+i0.25111$ & $ 10.056 -i11.669 $ & 2.357 \\
\cline{2-4}
$\eta\,{^3\!He}$ & $-0.51111+i0.30709$ & $ 7.0788 -i13.312 $ & 3.316 \\
\cline{2-4}
    & $-0.47578+i0.34354$ & $ 4.5944 -i13.863 $ & 7.617 \\
\hline\hline
    & $-0.15056+i0.18278$ & $-0.43713-i2.2399 $ & 2.357 \\
\cline{2-4}
$\phantom{-}\eta\,{^4\!He}\phantom{-}$ & $-0.17940+i0.24300$ &
                              $-1.0933 -i3.5484 $ & 3.316 \\
\cline{2-4}
    & $\phantom{-}-0.23100+i0.30850\phantom{-}$ &
      $\phantom{-}-1.7016-i5.8006\phantom{-}$ & 7.617 \\
\hline\hline
\end{tabular}
\end{center}
\vspace{5mm}
Table 2.\,\,
The enhancing factors $g$ moving the $\eta$--nucleus amplitude poles
to the points
$p_0=\sqrt{2\mu E_0}$ on the diagonal.
For each of the nuclei the calculations were done
with three values of the range parameter $\alpha$.
\begin{center}
\begin{tabular}{|c|c|c|c|c|}
\hline
 & $g$ & $p_0\;({\rm fm}^{-1})_{\mathstrut}^{\mathstrut}$ &
             $E_0\;({\rm MeV})$ &
             $\phantom{-}\alpha\;({\rm fm}^{-1})\phantom{-}$\\
\hline\hline
  & 1.654  & $-0.32545+i0.32545$ & $-i9.7134$ & 2.357 \\
\cline{2-5}
$\eta\, d$ & 1.566 & $-0.33741+i0.33741$ & $-i10.440$ & 3.316 \\
\cline{2-5}
    & 1.535 & $-0.33938+i0.33938$ & $-i10.566$ & 7.617 \\
\hline\hline
    & 1.361 & $-0.33900+i0.33900$ & $-i9.7467$ & 2.357 \\
\cline{2-5}
$\eta\, t$ & 1.310 & $-0.35424+i0.35424$ & $-i10.643$ & 3.316 \\
\cline{2-5}
    & 1.260 & $-0.35378+i0.35378$ & $-i10.615 $ & 7.617 \\
\hline\hline
    & 1.330 & $-0.34375+i0.34375$ & $-i10.022$ & 2.357 \\
\cline{2-5}
$\eta\,{^3\!He}$ & 1.221 & $-0.36640+i0.36640$ & $-i11.386$ & 3.316 \\
\cline{2-5}
    & 1.144 & $-0.38004+i0.38004$ & $-i12.247$ & 7.617 \\
\hline\hline
    & 0.955 & $-0.16164+i0.16164$ & $-i2.1267 $ & 2.357 \\
\cline{2-5}
$\phantom{-}\eta\,{^4\!He}\phantom{-}$ & 0.911 & $-0.19940+i0.19940$ &
                              $-i3.2363 $ & 3.316 \\
\cline{2-5}
    & $\phantom{-}0.899\phantom{-}$ &
      $\phantom{-}-0.25130+i0.25130\phantom{-}$ &
      $\phantom{-}-i5.1403\phantom{-}$ & 7.617 \\
\hline\hline
\end{tabular}
\end{center}

These factors correspond to an  $\eta N$ attraction, which just generates
an $\eta$--nucleus binding with $Re\,E_0=0$. Further increase of
$g$ moves the poles up and to the right, enhancing the binding and reducing
the widths of the states.\\

It is seen, that the real part of the $a_{\eta N}$, providing the critical
binding of $\eta$--meson to a light nucleus, lies within the existing
uncertainties, $[0.27,0.98]$ \cite{serb}, for this value. Therefore, in
reality the $\eta$--nucleus quasi--bound states can exist with $A\ge2$.
If this is not the case, then at least the near--threshold resonances ( poles
just below the
diagonal ) must exist. However, as one sees from both  tables, the widths
of such quasi--bound and resonance states are small only for  the
$\eta{^4\!He}$ system while for the other systems considered, they
 are rather large $\sim20$ MeV, which  means that such states is difficult
to be detected in experiments.



\end{document}